\documentclass[11pt]{article}

\usepackage[preprint]{acl}

\usepackage{times}
\usepackage{latexsym}

\usepackage[T1]{fontenc}

\usepackage[utf8]{inputenc}

\usepackage{microtype}

\usepackage{inconsolata}
\usepackage{multirow}
\usepackage{hyperref}
\usepackage{comment}
\usepackage{color}
\usepackage{graphicx}
\usepackage{amsmath,amsfonts}
\usepackage{booktabs}
\usepackage{colortbl}

\usepackage{longtable}
\title{HAM-TTS: Hierarchical Acoustic Modeling for Token-Based Zero-Shot Text-to-Speech with Model and Data Scaling}

\author{
Chunhui Wang\footnotemark[1] \textsuperscript{1},
Chang Zeng\footnotemark[1] \textsuperscript{2 3},
Bowen Zhang\footnotemark[1] \textsuperscript{1},
Ziyang Ma\footnotemark[1] \textsuperscript{4}, \\
\textbf{
Yefan Zhu\textsuperscript{1},
Zifeng Cai\textsuperscript{1},
Jian Zhao\textsuperscript{1},
Zhonglin Jiang\textsuperscript{1},
Yong Chen\textsuperscript{1}} \\
$^1$ Geely, China, $^2$ National Institute of Informatics, Japan, \\
$^3$ SOKENDAI, Japan, $^4$ Shanghai Jiao Tong University, China \\ 
\small{\texttt{\{Chunhui.Wang5,bowen.zhang3\}@geely.com, zengchang@nii.ac.jp, zym.22@sjtu.edu.cn}} \\
\small{\texttt{\{Yefan.Zhu, Zifeng.Cai, Jian.Zhao9, zhonglin.jiang, yong.chen\}@geely.com}}
}

\begin{document}

\maketitle
\renewcommand{\thefootnote}{\fnsymbol{footnote}}
\footnotetext[1]{These authors contributed equally to this work.}
\renewcommand{\thefootnote}{\arabic{footnote}}

\begin{abstract}
Token-based text-to-speech (TTS) models have emerged as a promising avenue for generating natural and realistic speech, yet they grapple with low pronunciation accuracy, speaking style and timbre inconsistency, and a substantial need for diverse training data. In response, we introduce a novel hierarchical acoustic modeling approach complemented by a tailored data augmentation strategy and train it on the combination of real and synthetic data, scaling the data size up to 650k hours, leading to the zero-shot TTS model with 0.8B parameters. Specifically, our method incorporates a latent variable sequence containing supplementary acoustic information based on refined self-supervised learning (SSL) discrete units into the TTS model by a predictor. This significantly mitigates pronunciation errors and style mutations in synthesized speech. During training, we strategically replace and duplicate segments of the data to enhance timbre uniformity. Moreover, a pretrained few-shot voice conversion model is utilized to generate a plethora of voices with identical content yet varied timbres. This facilitates the explicit learning of utterance-level one-to-many mappings, enriching speech diversity and also ensuring consistency in timbre. Comparative experiments\footnote{Demo page: \href{https://anonymous.4open.science/w/ham-tts/}{https://anonymous.4open.science/w/ham-tts/}} demonstrate our model's superiority over VALL-E in pronunciation precision and maintaining speaking style, as well as timbre continuity.
\end{abstract}

\section{Introduction}
\label{sec:introduction}

In the last decade, significant strides \cite{gan,vae,vq-vae,flow,transformer,diffusion} have been made in the advancement of deep learning and neural network technologies, enabling the text-to-speech (TTS) to evolve from the cascade manner of acoustic models \cite{tacotron,transformertts,glow-tts,grad-tts} and vocoders \cite{wavenet,hifigan,hifiwavegan,diffwave} to the fully end-to-end (E2E) style \cite{fs2,vits,valle,megatts2,tts-survey}. These methods are not only capable of rapidly generating high-quality speech, but also adept at synthesizing more challenging vocal expressions such as singing \cite{xiaoice,xiaoice2,crosssinger}. However, most TTS systems utilize continuous acoustic features such as MFCC in the frequency domain as intermediate representations for modeling, hindering from generating high-quality speech in the zero-shot scenario of timbre due to their mixture of semantic and acoustic information and difficulty of disentanglement \cite{speechgpt,hificodec}.

Recently, token-based TTS \cite{audiolm,valle,instructtts,naturalspeech2,wang2023lauragpt,song2024ella} methods have attracted extensive attention from both academia and industry due to their potential for synthesizing high-quality speech in the zero-shot scenario. Among these, the neural audio codec \cite{soundstream,encodec,hificodec} has demonstrated immense potential to serve as the intermediate representation for TTS modeling. For example, VALL-E \cite{valle} utilizes a large language model \cite{gpt2,gpt3,llama,llama2} to approximate the distribution of neural audio codecs \cite{encodec} and can synthesize speech that closely mimics a target speaker's voice from a mere three-second sample. However, despite their promising capabilities, we observe that these models often struggle with maintaining accurate pronunciation and consistent speaking style as well as timbre in synthesized speech. Additionally, the substantial requirement for large and diverse training data further limits their widespread adoption.

To tackle these issues, we proposed a \textbf{H}ierarchical \textbf{A}coustic \textbf{M}odeling method, namely HAM-TTS, with a tailored data augmentation strategy for the token-based TTS model \cite{audiolm,valle,instructtts}. Specifically, in order to alleviate the difficulty of directly modeling the mapping from text to neural audio codec in previous studies, we incorporate a latent variable sequence (LVS) containing supplementary acoustic information based on HuBERT \cite{hubert} features into the TTS model. A Text-to-LVS predictor is optimized simultaneously with TTS model. In the inference stage, the text prompt is converted to the LVS by the predictor to provide imperative acoustic information to mitigate pronunciation errors.

Unfortunately, generating LVS based on simple HuBERT features cannot revise the issue of inconsistency of speaking style in the synthesized speech due to the personalized information contained in HuBERT features, which is a distractor to the audio prompt. Therefore, we applied the K-Means \cite{kmeans} clustering method to refine HuBERT features for removing personalized information such as speaking styles, enabling the TTS model to make use of the remaining acoustic information to improve pronunciation accuracy while maintaining consistent speaking style with the audio prompt throughout the entire synthesized speech.

Timbre inconsistency is another serious problem for token-based TTS systems \cite{audiolm}. We designed a timbre consistency data augmentation strategy to train the proposed HAM-TTS system to revise it. Concretely, we randomly replace a successive segment of a training sample with a small chunk selected from other training utterances or duplicate a successive segment of a training sample while forcing the model to predict the original utterance. It enhances the timbre consistency of the synthesized speech in the zero-shot scenario.

As illustrated in \cite{audiolm,valle,naturalspeech2}, token-based TTS methods require extensive training data to assign the model the ability to synthesize diverse and high-quality speech. In this paper, instead of solely using substantial real speech data for training, we utilized a pretrained UNet-based \cite{unet} few-shot voice conversion model to generate voices with the same content but different timbres as a supplementary dataset, enabling the model to explicitly learn one-to-many mapping knowledge, which is beneficial to improve the diversity of generated speech and the timbre consistency.

We trained many models with different configurations on a large-scale internal Chinese dataset and evaluated them on the public AISHELL1 dataset \cite{aishell1}. We rigorously compared HAM-TTS against the state-of-the-art (SOTA) VALL-E model, which served as our baseline. The results of these experiments, conducted on a substantial dataset, clearly establish the advantages of our approach over the baseline model, demonstrating the enhanced capabilities of HAM-TTS, particularly in terms of pronunciation accuracy, speaking style consistency, and timbre continuity in challenging zero-shot scenarios.

This paper is structured to provide a comprehensive overview of our research and findings. Following this introduction, some related works are introduced in Section \ref{sec:related}. We delve into the specifics of our hierarchical acoustic modeling method in Section \ref{sec:hamtts}. We then present the experimental setup and results, offering a comparative analysis with current benchmarks in Section \ref{sec:exp}. The paper concludes with a summary of our contributions and a discussion on future research directions in Section \ref{sec:con}.

\section{Related Works}
\label{sec:related}
Although there are many studies \cite{hts,rnntts,tacotron,transformertts,fs2,vits,valle} focusing on TTS, in this section, we only briefly review some representative works about neural audio codecs and speech generative models based on them for a closer connection to our work.

\subsection{Neural Audio Codec}

Recent advancements in neural audio codecs, as illustrated in \cite{soundstream,encodec,hificodec}, have significantly enhanced the field of speech synthesis. These studies collectively highlight the efficiency of neural codecs in encoding and decoding audio data, offering a more compact and flexible representation compared to traditional methods.

Soundstream \cite{soundstream} introduces a novel end-to-end neural audio codec framework, demonstrating effective compression of audio signals into a discrete latent space by residual vector quantization. This advancement facilitates the generation of high-quality audio from compact representations, highlighting the codec's versatility in various audio applications.

Encodec \cite{encodec} further explores this domain, emphasizing the codec's role in efficiently compressing audio data while maintaining quality. Its approach showcases the potential of neural codecs in handling complex audio tasks with reduced data requirements, a crucial factor in resource-constrained environments.


In our research, these insights into neural audio codecs lay the foundation for developing a robust and efficient token-based TTS model. The enhanced fidelity and efficiency of neural codecs directly inform our approach, enabling us to achieve superior speech synthesis quality, particularly in zero-shot scenarios.

\begin{figure*}[t]
    \centering
    \includegraphics[width=0.7\linewidth]{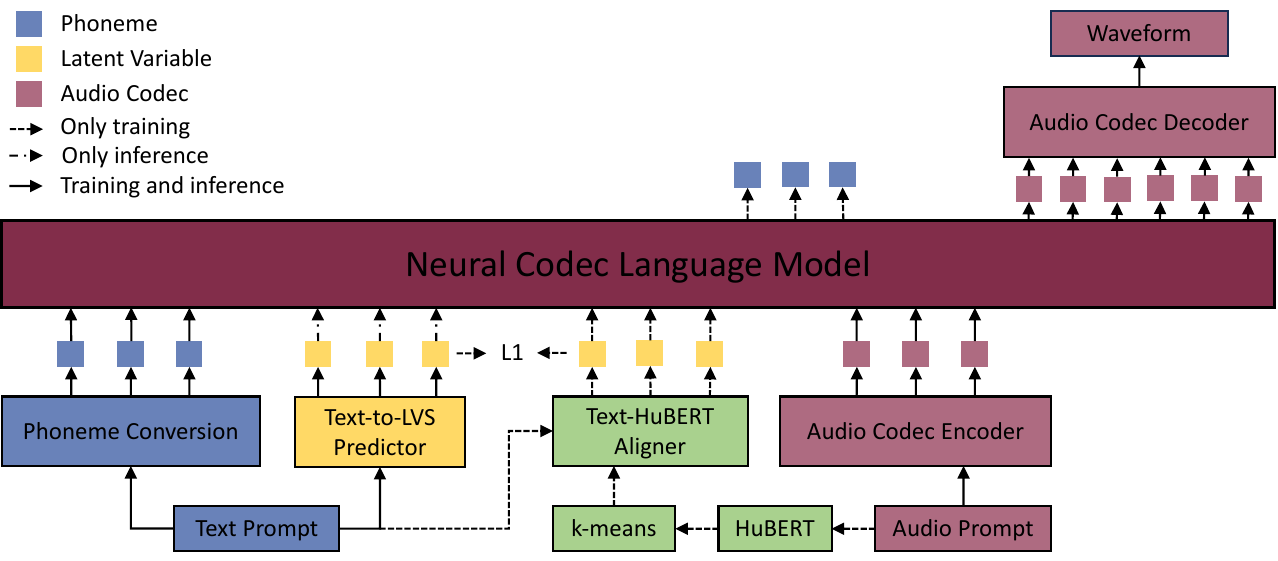}
    \caption{Overview of HAM-TTS. Although it builds upon VALL-E, its design including Text-HuBERT aligner and Text-to-LVS is applicable across various token-based TTS models. To enhance the ability of HAM-TTS to process semantic information, we also let codec language models predict the phoneme sequence based on the input text in the training stage.}
    \label{fig:overview-HAMTTS}
    \vspace{-4mm}
\end{figure*}

\subsection{Token-based Speech Generation Model}
More and more studies \cite{audiolm,valle,naturalspeech2,wang2023lauragpt,song2024ella} are beginning to try to use neural audio codecs as intermediate representations for speech generation. These approaches highlight the growing consensus in the field regarding the effectiveness of neural codecs in handling complex tasks.

AudioLM \cite{audiolm} represents a significant leap in audio generation by employing a language modeling approach. It particularly stands out for its ability to generate coherent and contextually appropriate speech, attributed to its advanced use of latent vectors conditioned on inputs. This model demonstrates how the integration of neural codecs \cite{soundstream} can facilitate the production of diverse and high-quality speech.

VALL-E \cite{valle}, on the other hand, capitalizes on the neural codec's ability \cite{encodec} to approximate large language models, enabling the synthesis of speech that closely mimics a target speaker's voice from a minimal sample. 

NaturalSpeech2 \cite{naturalspeech2} takes these concepts further by integrating a neural audio codec with additional components such as the diffusion model. Its emphasis on zero-shot synthesis capabilities and prosody highlights the model's robustness and versatility, particularly in generating diverse speech styles and maintaining voice quality across various scenarios.

These studies collectively underscore the importance of neural codecs in speech generation and pave the way for our research. In our work, we build upon these foundations and propose a novel hierarchical acoustic modeling approach to enhance pronunciation accuracy and speaking style consistency while utilizing a data augmentation strategy and synthetic data to emphasize the timbre consistency and diversity of generated voices.

\section{HAM-TTS}
\label{sec:hamtts}

The introduction of the HAM-TTS model is presented in this section. As depicted in Figure \ref{fig:overview-HAMTTS}, 
in addition to the phoneme conversion and audio codec encoder components originating from the existing TTS model like VALL-E, we design a predictor to directly transform the text prompt to the latent variable sequence (LVS) to incorporate supplementary acoustic information into the neural codec language model in the inference stage. The predictor is jointly optimized with the TTS model in the training stage via the supervising signal from the output of the Text-HuBERT aligner, which utilizes the cross-attention mechanism \cite{blip2} to align the phoneme sequence and the HuBERT features refined by K-Means clustering to generate the LVS. Detailed designs of the Text-HuBERT aligner and the Text-to-LVS predictor are presented in Section \ref{sec:ham}. The timbre consistency data augmentation strategy is another important contribution of our work for revising the issue of timbre inconsistency in synthesized speech. It is concretely illustrated in Section \ref{sec:dataaug}. Finally, the supplementary synthetic dataset generated by the pretrained few-shot voice conversion model is elaborated in Section \ref{sec:synthetic}. Detailed configurations for models used in our experiment will be illustrated in Appendix \ref{app:config}.

\begin{figure}[t]
    \centering
    \includegraphics[width=0.25\linewidth]{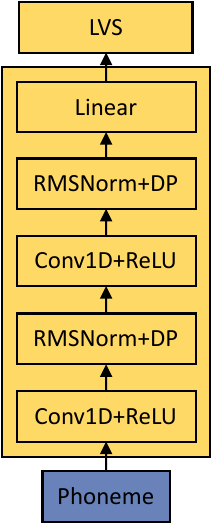}
    \caption{Structure of Text-to-LVS predictor. ``DP'' means dropout \cite{dropout} operation. It learns the mapping from the text prompt to the LVS in the training stage. Once the training is complete, it can generate the LVS from the text prompt directly in the inference stage.}
    \label{fig:text2lvs}
    \vspace{-4mm}
\end{figure}

\subsection{Hierarchical Acoustic Modeling}
\label{sec:ham}
We observed that previous studies like AudioLM \cite{audiolm} and VALL-E \cite{valle} occasionally produced speech with incorrect pronunciation. This was largely due to the limitations in directly mapping text to a neural audio codec sequence without adequate acoustic information. To address this, the Text-to-LVS predictor shown in Figure \ref{fig:text2lvs} is proposed to generate the latent variable sequence containing the imperative acoustic information from the phoneme sequence in the inference stage, which can be formulated as,
\begin{equation}
\label{eq:lvs-inference}
    \boldsymbol{L}'_{1:T_1} = f_{pred}(\boldsymbol{X}_{1:T_1}),
\end{equation}
where $\boldsymbol{X}_{1:T_1}$ represents the phoneme sequence with $T_1$ phoneme units. $f_{pred}(\cdot)$ denotes the function of the predictor's transformation. $\boldsymbol{L}'_{1:T_1}$ is the generated LVS with the same length of the phoneme sequence. The LVS is concatenated with the corresponding phoneme sequence. Following this concatenation, the combined sequence is transformed via a convolutional layer to align with the dimension required by the neural audio codec before feeding them to the codec language model. It can be represented as,
\begin{align}
    \boldsymbol{S}_{1:T_1} = \text{Conv1d}(\text{Concat}(\boldsymbol{X}_{1:T_1},\boldsymbol{L}'_{1:T_1})),
\end{align}
where $\boldsymbol{S}_{1:T_1}$ is the output aligning with the dimension of audio codecs.

\begin{figure}[t]
    \centering
    \includegraphics[width=0.65\linewidth]{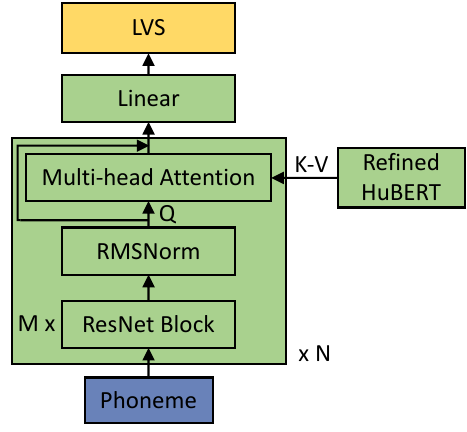}
    \caption{Structure of Text-HuBERT aligner. It utilizes the text prompt and the refined HuBERT feature as input to generate the LVS in the training stage. The generated LVS is also used as a supervising signal to train the Text-to-LVS predictor.}
    \label{fig:aligner}
    \vspace{-4mm}
\end{figure}

As illustrated in Figure \ref{fig:overview-HAMTTS}, the Text-to-LVS predictor is simultaneously optimized with the neural codec language model in the training stage via the supervising signal generated from another new module, namely Text-HuBERT aligner. The aligner consists of $N$ blocks with the same architecture as shown in Figure \ref{fig:aligner}. Each block contains $M$ residual convolution networks \cite{resnet} denoted as ResNet Block in the figure, followed by a root mean square layer normalization (RMSNorm) \cite{rmsnorm}, and finally a multi-head attention layer \cite{transformer} is utilized to align the output sequence of RMSNorm with the HuBERT \cite{hubert} features (key and value) refined by K-Means clustering. Unlike the standard layer normalization used in the Transformer model \cite{transformer}, we employ RMSNorm in the aligner, enhancing its capability to handle complex sequences and achieve faster convergence. The supervising LVS with the same length of the phoneme sequence can be computed by,
\begin{equation}
\label{eq:lvs-training}
    \boldsymbol{L}_{1:T_1} = f_{aligner}(\boldsymbol{X}_{1:T_1}, \boldsymbol{H}_{1:T_2}),
\end{equation}
where $\boldsymbol{H}_{1:T_2}$ is the refined HuBERT feature sequence with $T_2$ length and $\boldsymbol{L}_{1:T_1}$ denotes the supervising LVS. $f_{aligner}(\cdot)$ means the function of Text-HuBERT aligner module. Note that it is imperative to leverage the K-Means clustering to remove personalized information from the original HuBERT feature for revising the mutation of speaking style in synthesized speech in the zero-shot scenario.

The approximation between $\boldsymbol{L}_{1:T_1}$ and $\boldsymbol{L}'_{1:T_1}$ is measured by a L1 loss function shown as,
\begin{equation}
\label{eq:lvs-loss}
    \mathcal{L}_{LVS} = \sum_{t=1}^{T_1}|\boldsymbol{L}'_t - \boldsymbol{L}_t|,
\end{equation}
where $\mathcal{L}_{LVS}$ is the metric measuring how close the $\boldsymbol{L}'_{1:T_1}$ is to $\boldsymbol{L}_{1:T_1}$.

\subsection{Timbre Consistency Data Augmentation}
\label{sec:dataaug}

Timbre inconsistency of the synthesized speech has been a non-negligible problem plaguing the TTS system in the zero-shot scenario despite the fact that contemporary token-based TTS systems \cite{valle,instructtts} claim to enable timbre cloning. In this section, we will illustrate our proposed timbre consistency data augmentation strategy for this issue.

To ensure timbre consistency in the synthesized speech, we implemented a data augmentation strategy on our training data. Specifically, during the loading of a batch of speech data, for 10\% portion, we either randomly select a continuous segment from another sample to replace a segment in the current sample, or we randomly duplicate a segment from the same sample and concatenate it to the end of that segment. In the loss calculation, neural audio codecs from samples without data augmentation are treated as ground truth for computing the cross-entropy loss with the generated codecs. This approach enables the model to develop strong resistance to timbre perturbations. Consequently, it prevents short-term timbre variations from affecting the timbre of the entire generated speech segment, thus ensuring timbre consistency in the synthesized speech.

\subsection{Supplmentary Synthetic Dataset}
\label{sec:synthetic}

The fact that extensive speech data are needed to train a TTS model is prohibitive for many academic researchers. For example, Audiobox \cite{audiobox} has scaled the size of the training data up to 100k hours, which is a heavy burden to collect that much data for academic institutions. At the same time, there are many legal risks associated with using real data without authorization. These facts motivate us to consider using synthetic data to train TTS models. In this section, we will show how to generate synthetic data as a supplementary dataset for real data.

%
It is difficult to collect a large amount of data for voices with single timbre and long duration in the real world, especially for more than ten seconds, which leads to sparse data for speech with long duration when training speech synthesis models and also makes it more difficult for the model to ensure the consistency of the timbre of the whole sentence when generating long speech. With this in mind, we utilize a pre-trained UNet-based \cite{unet} few-shot voice conversion model concretely illustrated in Appendix \ref{app:vc} to generate a large amount of long speech data to compensate for the lack of real data. We randomly select 1,000 speakers with a few minutes of speech from the real data as candidates and convert around 500 hours of real speech whose duration ranges from 10 to 20 seconds in the training dataset for each candidate. Consequently, the large amount of synthetic data improves the diversity of training data by explicitly providing one-to-many mapping for the scenario of long voices, distinct from previous studies \cite{tacotron,fs2,audiolm,valle} in which only the phoneme-level diversity was considered.

\subsection{Loss Function}
\label{sec:loss}
We follow the training strategy of VALL-E \cite{valle} regarding TTS as a conditional codec language modeling task. Two Transformer \cite{transformer} decoder-only codec language models are trained for autoregressive (AR) and non-autoregressive (NAR) modeling, respectively. We utilize the cross-entropy (CE) loss function to measure the distance between the real and the learned distribution of codecs. It can be formulated as,
\begin{align}
    \mathcal{L}_{codecs} & = \sum_{t=1}^{T_3}\text{CE}(\boldsymbol{A}_t,\boldsymbol{A}'_t),
\end{align}
where $\boldsymbol{A}$ and $\boldsymbol{A}'$ mean codec sequences of the ground truth and synthesized one, respectively. $T_3$ denotes the length of the codec sequence. $\mathcal{L}_{codecs}$ is the loss for codec generation.

Moreover, to enhance the ability of HAM-TTS to process semantic information, the teacher forcing loss is computed on the AR codec LM and the NAR codec LM to fit the distribution of input texts, and the corresponding CE loss function is shown as,
\begin{align}
    \mathcal{L}_{phoneme} & = \sum_{t=1}^{T_1}\text{CE}(\boldsymbol{X}_t,\boldsymbol{X}'_t),
\end{align}
where $\boldsymbol{X}'$ means the synthesized phoneme sequence. $\mathcal{L}_{phoneme}$ is the loss for text generation.

The total loss is the sum of three loss terms, illustrated as Eq. \ref{eq:loss}. More details of the training method are available in Appendix \ref{app:training}.
\begin{equation}
\label{eq:loss}
    \mathcal{L} = \mathcal{L}_{LVS} + \mathcal{L}_{phoneme} + \mathcal{L}_{codecs}
\end{equation}


\section{Experiment}
\label{sec:exp}

\begin{table*}
  \caption{Performance comparison on AISHELL1 dataset. All models were trained exclusively on 150k hours of real data. We compare the performance of ground truth (GT), VALL-E, HAM-TTS-S, and HAM-TTS-L models, showcasing the effectiveness of HAM-TTS in pronunciation accuracy, naturalness, and speaker similarity. The NMOS, SMOS, and MOS were computed with a 95\% confidence interval.}
  \centering
  \small
  \label{tab:primary_exp}
  \setlength\tabcolsep{5pt}
  \begin{tabular}{lrrrrr}
    \toprule
    \textbf{Model} & \textbf{\#Params} & \textbf{CER}\%($\downarrow$) & \textbf{NMOS}($\uparrow$) & \textbf{SMOS}($\uparrow$) & \textbf{MOS}($\uparrow$) \\
    \midrule
    GT & - & 2.6 & 4.03$\pm$0.08 & 4.30$\pm$0.06 & 4.45$\pm$0.07 \\
    \midrule 
    VALL-E & 426M & 5.5 & 3.65$\pm$0.15 & 4.03$\pm$0.12 & 4.05$\pm$0.10 \\
    HAM-TTS-S & 421M & 4.0 & 3.79$\pm$0.11 & 4.12$\pm$0.10 & 4.27$\pm$0.08 \\
    \rowcolor{lightgray}
    HAM-TTS-L & 827M & 3.2 & 4.01$\pm$0.07 & 4.26$\pm$0.09 & 4.45$\pm$0.07 \\
    \bottomrule
  \end{tabular}
\end{table*}

\subsection{Experiment Setup}
\textbf{Dataset}: All TTS models were trained on our internal Chinese speech dataset comprising both real and synthetic speech. The dataset includes 150k hours of real speech and 500k hours of synthetic speech. The real speech component encompasses approximately 20,000 speakers, with each audio segment ranging between 5 to 20 seconds in length and a sampling rate of 24kHz. On the other hand, the synthetic speech dataset is derived from 1,000 speakers, with each audio segment varying from 10 to 20 seconds in length. This extensive and diverse dataset plays a critical role in the robust training and performance of our model. As for the test data, we selected 50 speakers from the public AISHELL1 dataset \cite{aishell1} and each speaker has five sentences whose duration varies from 5-20 seconds. Since our training data has no overlap with the public dataset, all testing speakers are unseen, aiming at showing the zero-shot ability of our model.

\noindent\textbf{Baseline}: VALL-E \cite{valle} is used as the baseline model in our experiments since it is a representative SOTA work of token-based TTS systems. We reproduced and trained it on the internal dataset due to no official implementation available.

\noindent\textbf{Evaluation metrics}: We evaluate all models from three aspects: pronunciation accuracy, speaking style consistency, and timbre consistency. Pronunciation accuracy is represented by character error rate (CER) metric, which is calculated by a pretrained Whisper \cite{whisper} model\footnote{\href{https://huggingface.co/espnet/pengcheng_aishell_asr_train_asr_whisper_medium_finetune_raw_zh_whisper_multilingual_sp}{https://huggingface.co/espnet/pengcheng\_aishell\_asr\_tra\\in\_asr\_whisper\_medium\_finetune\_raw\_zh\_whisper\_multilin\\gual\_sp}} provided by ESPNet \cite{espnet}. Speaking style consistency is evaluated by mean opinion score regarding the naturalness (NMOS) of speech since the mutation of speaking style is perceptible from the feedback of listeners. Timbre consistency is evaluated by the speaker similarity MOS (SMOS) metric. Additionally, we also requested all listeners to evaluate the overall quality of testing data, including the naturalness, audio quality, and pronunciation accuracy. It is represented as the MOS metric. As for the number of listeners, we employed 60 people to participate in the test. Each listener will evaluate the performance for all utterances. We believe that a listening test of this magnitude would provide a relatively objective result for the experiment.

\subsection{Primary Experimental Result}

In our experimental analysis, as detailed in Table \ref{tab:primary_exp}, all models were trained exclusively on 150k hours of real data. The HAM-TTS model, designed in two variants, HAM-TTS-S and HAM-TTS-L, explores different scales of parameterization. HAM-TTS-S, matching VALL-E with 421M parameters, ensures a fair comparison, while HAM-TTS-L expands to 827M parameters, aiming to unlock the full potential of the HAM-TTS. This scaling is crucial for assessing the effectiveness of our model in various parameter configurations.

In the table, our reproduced VALL-E achieves a CER of 5.5\%, an NMOS of 3.65, an SMOS of 4.03, and an overall MOS of 4.05, aligning with those presented in the original VALL-E paper \cite{valle}, indicating the reliability of our experimental setup. These results demonstrate VALL-E's proficiency in generating speech, but also highlight areas for improvement, particularly in pronunciation accuracy and naturalness compared with the result of GT. The HAM-TTS-S model achieves a CER of 4.0\%, lower than VALL-E's 5.5\%, indicating better pronunciation accuracy. Its NMOS at 3.79 and SMOS at 4.12 also surpass VALL-E, suggesting improved perceived quality and speaker similarity. The HAM-TTS-L further improves these metrics, recording a CER of 3.2\%, and comparable NMOS and SMOS scores to GT, illustrating the scalability and effectiveness of the HAM-TTS model in generating high-quality, realistic speech. These results demonstrate the HAM-TTS model's superiority in pronunciation accuracy and the consistency of speaking style and timbre.

\subsection{Ablation Study of K-Means}

\begin{table}
  \small
  \caption{A comparison of HAM-TTS-S model performance with and without K-Means clustering is provided to highlight the improvement in CER, NMOS, and overall MOS metrics due to K-Means feature refinement.}
  \centering
  \label{tab:kmeans}
  \setlength\tabcolsep{5pt}
  \begin{tabular}{lrrr}
    \toprule
    \textbf{Model} & \textbf{CER}\%($\downarrow$) & \textbf{NMOS}($\uparrow$) & \textbf{MOS}($\uparrow$) \\
    \midrule
    GT & 2.6 & 4.30$\pm$0.06 & 4.45$\pm$0.09 \\
    \midrule 
    w/o K-Means & 4.2 & 3.63$\pm$0.12 & 4.14$\pm$0.08 \\
    \rowcolor{lightgray}
    HAM-TTS-S & 4.0 & 3.79$\pm$0.11 & 4.27$\pm$0.08 \\
    \bottomrule
  \end{tabular}
  \vspace{-4mm}
\end{table}

In our HAM-TTS model, we employed the K-Means clustering technique to refine HuBERT features. This approach aims to remove personalized information such as speaking styles, enabling the TTS model to focus on the core acoustic information for enhancing pronunciation accuracy and maintaining consistent speaking style with the audio prompt throughout the synthesized speech.

Table \ref{tab:kmeans} in our experimental results presents the effectiveness of the K-Means clustering in our model. We compared the performance of HAM-TTS-S with and without the application of K-Means clustering. The results demonstrate that the application of K-Means clustering further improves the model's performance. Specifically, the CER for the HAM-TTS-S without K-Means clustering was 4.2\%, while the implementation of K-Means clustering reduced the CER to 4.0\%. This reduction in CER indicates an improvement in pronunciation accuracy, which is a direct result of the refined HuBERT features providing more accurate acoustic information.

Furthermore, the NMOS and the overall MOS also slightly improved with the use of K-Means clustering. The NMOS increased from 3.63 to 3.79, and the MOS increased from 4.14 to 4.27, indicating that the speech synthesized with the refined features was perceived as more natural and of higher quality by listeners. These results clearly illustrate the impact of K-Means clustering in enhancing the overall performance of the HAM-TTS-S, affirming its effectiveness in providing a more accurate and consistent speaking style in synthesized speech.

\subsection{Ablation Study of Synthetic Data}

\begin{table}
  \small
  \caption{Experimental result to show the effectiveness of synthetic data. We trained the HAM-TTS-S model with different sizes and combinations of real(R) and synthetic(S) data.}
  \centering
  \label{tab:synthetic}
  \setlength\tabcolsep{3pt}
  \begin{tabular}{lrrr}
    \toprule
    \textbf{Training data} & \textbf{CER}\%($\downarrow$) & \textbf{SMOS}($\uparrow$) & \textbf{MOS}($\uparrow$) \\
    \midrule
    GT & 2.6 & 4.30$\pm$0.06 & 4.45$\pm$0.07 \\
    \midrule 
    150k(R) & 4.0 & 4.12$\pm$0.10 & 4.27$\pm$0.08  \\
    150k(R)+150k(S) & 3.6 & 4.26$\pm$0.09 & 4.32$\pm$0.07 \\
    \rowcolor{lightgray}
    150k(R)+500k(S) & 2.8 & 4.32$\pm$0.07 & 4.49$\pm$0.08 \\
    \midrule
    150k(S) & 4.5 & 4.05$\pm$0.10 & 4.10$\pm$0.13 \\
    300k(S) & 4.1 & 4.13$\pm$0.07 & 4.25$\pm$0.08 \\
    \rowcolor{lightgray}
    500k(S) & 3.3 & 4.25$\pm$0.06 & 4.35$\pm$0.06 \\
    \bottomrule
  \end{tabular}
  \vspace{-4mm}
\end{table}

In our HAM-TTS model, synthetic data plays a pivotal role in enhancing the diversity and quality of the generated speech. We focused on demonstrating the impact of this synthetic data through a series of experiments, the results of which are detailed in Table \ref{tab:synthetic}.

The experiments were conducted using the HAM-TTS-S model, trained on different combinations and sizes of real and synthetic data. Our findings clearly show the significant improvements synthetic data brings to the model’s performance. When trained solely on 150k hours of real data, the HAM-TTS-S model achieves a CER of 4.0\%, an SMOS of 4.12, and an overall MOS of 4.27. However, when augmented with synthetic data, there is a marked improvement in all metrics. 

Specifically, training with an additional 150k hours of synthetic data (150k(R)+150k(S)) reduces the CER to 3.6\%, and further increases the SMOS to 4.26 and the MOS to 4.32. This improvement is even more pronounced when the model is trained with an additional 500k hours of synthetic data (150k(R)+500k(S)), resulting in a CER of 2.8\%, an SMOS of 4.32, and an MOS of 4.49. These results clearly indicate that synthetic data not only contributes to the reduction in pronunciation errors but also significantly enhances the quality of the synthesized speech since it enables the model to explicitly learn the knowledge of utterance-level one-to-many mappings.

Furthermore, the results underscore the promise of training HAM-TTS models solely on synthetic data. When the model was trained with varying amounts of synthetic data (150k(S), 300k(S), and 500k(S)), we observed a steady improvement in all evaluation metrics, approaching the performance levels of the model trained on real data. The model trained with 500k hours of synthetic data achieved a CER of 3.3\%, closely matching the 2.8\% CER of the model trained with a combination of real and synthetic data. This finding is particularly promising as it suggests that high-quality TTS systems can be developed even in scenarios where access to large amounts of real speech data is limited, highlighting the potential of synthetic data in training effective speech synthesis models.

These findings illustrate the significant impact of synthetic data in improving the performance of HAM-TTS models, both when used in conjunction with real data and when used exclusively, marking a substantial advancement in the field of speech synthesis.

\section{Conclusion and Future Work}
\label{sec:con}
In this study, we have introduced HAM-TTS, a novel text-to-speech system that leverages a hierarchical acoustic modeling approach. This system integrates advanced techniques such as K-Means clustering for refining HuBERT features and a comprehensive strategy incorporating both real and synthetic data. Our experiments demonstrate the effectiveness of HAM-TTS in improving pronunciation accuracy, speaking style consistency, and timbre consistency in zero-shot scenarios.

Despite these significant advancements, future work could explore the optimal combination of synthetic data in terms of speaker diversity and duration per speaker. This aspect could lead to further enhancements in handling a wide range of speech variations. Additionally, optimizing the inference speed of the HAM-TTS model is crucial for enhancing its practical usability, making it suitable for real-time applications and user interactions. The exploration of these avenues will contribute significantly to advancing the field of speech synthesis.

\section*{Limitation}
\label{sec:limit}
We acknowledge that while our HAM-TTS model has demonstrated significant advancements, certain aspects remain unexplored and present opportunities for future research. One such area is the optimal combination of synthetic data in terms of speaker diversity and duration per speaker. We have not yet investigated whether a greater number of speakers with less duration per speaker or fewer speakers but more duration per speaker would be more beneficial. This aspect is crucial for enhancing the model's ability to handle a wide range of speech variations and could potentially lead to further improvements in the model's performance.

Another limitation is the inference speed of the HAM-TTS model. Although the model achieves high-quality speech synthesis, the current inference process is not as efficient as it could be. There is considerable room for improvement in this area, particularly in terms of reducing the time taken to generate speech. Optimizing the model's architecture and streamlining the inference pipeline could significantly enhance the practical usability of HAM-TTS, making it more suitable for real-time applications and user interactions.

Addressing these limitations will be a focus of our future work, aiming to refine the HAM-TTS model further and expand its applicability in various speech synthesis scenarios.

\section*{Ethics Statement}
This research adheres to ethical standards in AI and speech synthesis, emphasizing data privacy, consent, and inclusivity. We address the potential for bias in our datasets and ensure fairness across diverse voices. Recognizing the risks of misuse, we advocate for responsible use and transparency in our methodology. Our work aims to contribute positively to technological advancements, balancing innovation with societal and individual well-being.

\bibliography{custom}


\appendix

\section{Appendix}
\label{sec:appendix}

\subsection{Model Details}
\label{app:config}

\begin{table*}
  \small
  \caption{Configuration of VALL-E in the experiment.}
  \centering
  \label{tab:valle}
  \setlength\tabcolsep{5pt}
  \begin{tabular}{ccc}
    \toprule
    \textbf{Component} & \textbf{Config} & \textbf{Value} \\
    \midrule
    Phoneme Conversion & Embedding Layer & 1024 \\
    \midrule 
    \multirow{2}{*}{Audio Codec Encoder\cite{encodec}} & Quantizer & 8 \\
    & Codebook Size & 1024 \\
    & Codebook Dimension & 1024 \\
    \midrule
    \multirow{5}{*}{Codec Language Model} & Attention Block & 14\footnotemark[3] \\
    & Heads & 16 \\
    & Hidden Size & 4096 \\
    & Dropout & 0.1 \\
    & Output Affine Layer & 1024 \\
    \bottomrule
  \end{tabular}
\end{table*}
\footnotetext[3]{In order to have a fair comparison with the HAM-TTS-S model, we increase the number of parameters of VALL-E to a comparable level by increasing two additional attention blocks.}

HAM-TTS is constructed based on the VALL-E framework, inheriting certain key architectural features. Similar to VALL-E, HAM-TTS incorporates two distinct Transformer decoders. These decoders are integral to the model's design, each serving a specific purpose in the speech synthesis process.

One of the Transformer decoders in HAM-TTS is dedicated to autoregressive modeling. This decoder plays a crucial role in sequentially predicting each element of the output based on the previously generated elements, thereby capturing the temporal dependencies in the speech sequence.

The other Transformer decoder in HAM-TTS is utilized for non-autoregressive modeling. This approach allows for the parallel generation of output elements, which can significantly enhance the model's efficiency by reducing the dependency on the sequential generation process.

Concrete configurations for VALL-E, HAM-TTS-S, and HAM-TTS-L are shown as Table \ref{tab:valle}, Table \ref{tab:ham-tts-s}, and Table \ref{tab:ham-tts-l}, respectively.

\begin{table*}
  \small
  \caption{Configuration of HAM-TTS-S in the experiment.}
  \centering
  \label{tab:ham-tts-s}
  \setlength\tabcolsep{5pt}
  \begin{tabular}{ccc}
    \toprule
    \textbf{Component} & \textbf{Config} & \textbf{Value} \\
    \midrule
    Phoneme Conversion & Embedding Layer & 1024 \\
    \midrule 
    \multirow{3}{*}{Audio Codec Encoder\cite{encodec}} & Quantizer & 8 \\
    & Codebook Size & 1024 \\
    & Codebook Dimension & 1024 \\
    \midrule
    \multirow{5}{*}{Codec Language Model} & Attention Block & 12 \\
    & Heads & 16 \\
    & Hidden Size & 4096 \\
    & Dropout & 0.1 \\
    & Output Affine Layer & 1024 \\
    \midrule
    \multirow{4}{*}{Text-to-LVS Predictor} & Conv1D Layers & 2 \\
    & Conv1D Kernel Size & 3 \\
    & Dropout & 0.1 \\
    & Output Affine Layer & 2 \\
    \midrule
    \multirow{8}{*}{Text-HuBERT Aligner} & Attention Block & 10 \\
    & Heads & 8 \\
    & Hidden Size & 4096 \\
    & Dropout & 0.1 \\
    & ResNet Block & 3 \\
    & Conv1D Layer & 2 \\
    & Conv1D Kernel Size & 3 \\
    & Output Affine Layer & 2 \\
    \bottomrule
  \end{tabular}
\end{table*}

\begin{table*}
  \small
  \caption{Configuration of HAM-TTS-L in the experiment.}
  \centering
  \label{tab:ham-tts-l}
  \setlength\tabcolsep{5pt}
  \begin{tabular}{ccc}
    \toprule
    \textbf{Component} & \textbf{Config} & \textbf{Value} \\
    \midrule
    Phoneme Conversion & Embedding Layer & 1024 \\
    \midrule 
    \multirow{3}{*}{Audio Codec Encoder\cite{encodec}} & Quantizer & 8 \\
    & Codebook Size & 1024 \\
    & Codebook Dimension & 1024 \\
    \midrule
    \multirow{5}{*}{Codec Language Model} & Attention Block & 24 \\
    & Heads & 16 \\
    & Hidden Size & 4096 \\
    & Dropout & 0.1 \\
    & Output Affine Layer & 1024 \\
    \midrule
    \multirow{4}{*}{Text-to-LVS Predictor} & Conv1D Layers & 2 \\
    & Conv1D Kernel Size & 3 \\
    & Dropout & 0.1 \\
    & Output Affine Layer & 2 \\
    \midrule
    \multirow{8}{*}{Text-HuBERT Aligner} & Attention Block & 10 \\
    & Heads & 8 \\
    & Hidden Size & 4096 \\
    & Dropout & 0.1 \\
    & ResNet Block & 3 \\
    & Conv1D Layer & 2 \\
    & Conv1D Kernel Size & 3 \\
    & Output Affine Layer & 2 \\
    \bottomrule
  \end{tabular}
\end{table*}

\subsection{Pretrained Voice Conversion Model}
\label{app:vc}
\begin{figure}[t]
    \centering
    \includegraphics[width=1\linewidth]{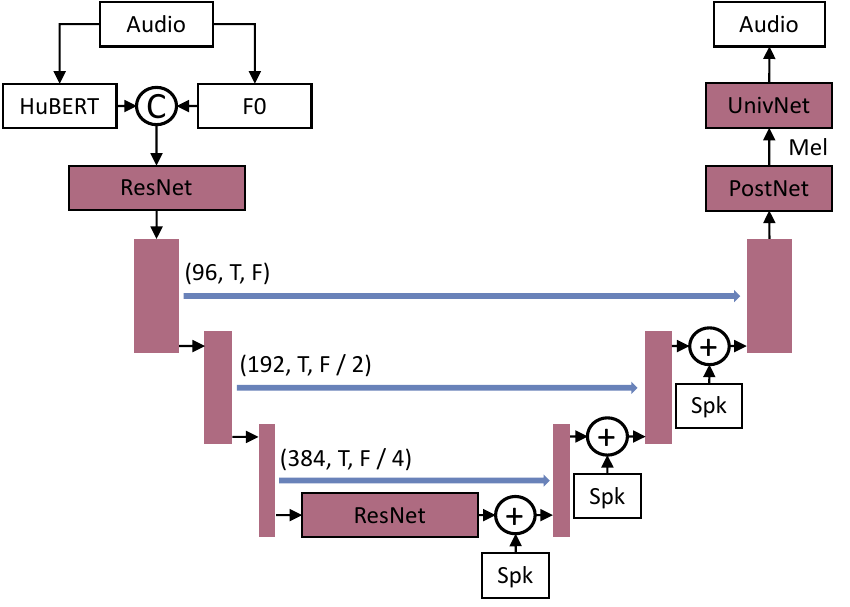}
    \caption{Structure of UNet-based voice conversion model. It is leveraged to generate extensive speech data with the same content but different timbres by several minutes of real speech from unseen target speakers.}
    \label{fig:unet}
\end{figure}

We employed a UNet-based \cite{unet} voice conversion model illustrated in Figure \ref{fig:unet} to generate 500k hours of speech data for training. 

In this voice conversion model, the initial processing stage involves extracting HuBERT and F0 features from the input audio. These extracted features are then concatenated and fed into a ResNet module for preprocessing. The ResNet module is designed to transform and refine these features, outputting them in the dimensions of (96, T, F), where `T' and `F' represent the time and frequency dimensions, respectively.

This output feature is then introduced into the encoder of a UNet architecture. The encoder performs downsampling on the frequency dimension twice, resulting in an output with dimensions (384, T, F / 4). Following this, another ResNet module is employed to further refine the output of the encoder. The refined features are then passed to the decoder of the UNet.

In the decoding process, the frequency dimension undergoes two stages of upsampling. Prior to each upsampling step, speaker characteristics are integrated into the input. This integration is crucial for ensuring that the synthesized speech retains the unique attributes of the speaker's voice. The final output from the decoder has the dimensions of (96, T, F), effectively restoring the original frequency dimension.

It is important to note that throughout the UNet architecture, the convolutional kernels used are of size (1,7). This specific kernel size aids in capturing the essential temporal and spectral characteristics of the speech signal.

The next stage involves the conversion of these processed features into the final waveform. This is achieved using a PostNet followed by a UnivNet vocoder \cite{univnet}, which together ensure the synthesized speech is both natural-sounding and closely matches the original audio in terms of timbre and prosody.

\subsection{Training Method}
\label{app:training}
We followed the training strategy used in VALL-E to employ a dual training approach to optimize the performance of the HAM-TTS model in both autoregressive (AR) and non-autoregressive (NAR) modeling. 

\noindent\textbf{AR Training}: The AR model is trained on the concatenation of the sequence $\boldsymbol{S}_{1:T_1}$ and the audio codec sequence $\boldsymbol{A}^{(1)}_{1:T_3}$ from the first quantizer of the Encodec model \cite{encodec}. It can be formulated as,
\begin{align}
& p(\boldsymbol{A}'^{(1)}|\boldsymbol{A}^{(1)},\boldsymbol{S};\theta_{AR}) = \\ \nonumber
& \prod_{t=0}^Tp(\boldsymbol{A}'^{(1)}_t|\boldsymbol{A}'^{(1)}_{<t},\boldsymbol{A}^{(1)},\boldsymbol{S};\theta_{AR})
\end{align}


\noindent\textbf{NAR Training}: The NAR model is employed for the audio codecs from the second to the last quantizers. This model is conditioned on $\boldsymbol{S}_{1:T_1}$, the acoustic prompt $\boldsymbol{A}^{(2:8)}_{1:T_3}$, and the predicted acoustic tokens $\boldsymbol{A}^{(<i)}_{1:T_3}$ from the previous codebooks. Each training step randomly samples a quantizer $i \in [2,8]$, and the model is trained to fit the distribution of codecs from the selected quantizer codebook. It can be formulated as,
\begin{align}
& p(\boldsymbol{A}'^{(2:8)}|\boldsymbol{A},\boldsymbol{S};\theta_{NAR}) \\
& =\prod_{i=2}^8p(\boldsymbol{A}'^{(i)}|\boldsymbol{A}'^{(<i)},\boldsymbol{A},\boldsymbol{S};\theta_{AR})
\end{align}

Both AR and NAR models were optimized using the Adam optimizer \cite{adam}, with a learning rate set at 0.03 and a warmup spanning the first 15,000 steps. After the warmup phase, the learning rate was managed using the CosineAnnealingLR scheduler \cite{cosine}. The training was conducted on a robust setup of 512 NVIDIA A100 80GB GPUs, and the model processed a batch size of 8k acoustic tokens. This extensive training was carried out over a total of 400k steps, leveraging the powerful computational capabilities of the A100 GPUs to efficiently handle the large batch size and extensive training steps.



\end{document}